\documentclass{article}

\usepackage{arxiv}
\usepackage{subcaption}
\usepackage[utf8]{inputenc} 
\usepackage[T1]{fontenc}    
\usepackage{hyperref}       
\usepackage{url}            
\usepackage{booktabs}       
\usepackage{amsfonts}       
\usepackage{nicefrac}       
\usepackage{microtype}      
\usepackage{lipsum}		
\usepackage{graphicx}
\usepackage{natbib}
\usepackage{doi}

\DeclareMathAlphabet{\mathcal}{OMS}{cmsy}{m}{n}
\newtheorem{mydef}{Definition}

\title{Uncertainty in fMRI Functional Networks of Autism Brain Imaging Data}

\author{ \href{https://orcid.org/0000-0001-5797-1068}{\includegraphics[scale=0.06]{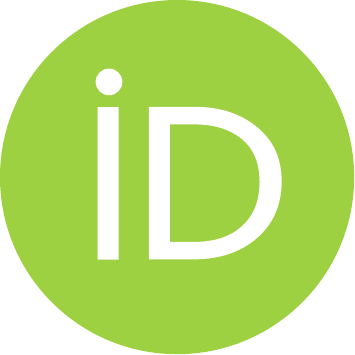}\hspace{1mm}Amin Kaveh}\thanks{Corresponding Author} \\
    Infolab\\
	Department of Information Technology\\
	Uppsala University\\
	75105 Uppsala, Sewden \\
	\texttt{amin.kaveh@it.uu.se} \\
	\And
	Matteo Magnani \\
	Infolab\\
	Department of Information Technology\\
	Uppsala University\\
	75105 Uppsala, Sewden \\
	\texttt{matteo.magnani@it.uu.se} \\
	\And
	Christian Rohner \\
	Infolab\\
	Department of Information Technology\\
	Uppsala University\\
	75105 Uppsala, Sewden \\
	\texttt{christian.rohner@it.uu.se} \\
}

\renewcommand{\shorttitle}{Uncertainty in fMRI Data}

\hypersetup{
pdftitle={Uncertainty in fMRI Functional Networks of Autism Brain Imaging Data},
pdfsubject={q-bio.NC, q-bio.QM},
pdfauthor={David S.~Hippocampus, Elias D.~Striatum},
pdfkeywords={First keyword, Second keyword, More},
}

\begin{document}
\maketitle



\section{Introduction} \label{sec:intro}
Graph analysis of functional neuroimaging data has attained extensive attention during the last decade. Among all functional imaging methods, resting state functional Magnetic Resonance Imaging (fMRI) has had a prominent role not only in identifying the intrinsic organization of the brain, but also in detecting the changes caused by psychiatric disorders. One way of analyzing fMRI data is to transform it into a brain network.

In the current approaches of modeling and analysis of resting state functional brain networks, similarity between different regions of the brain based on their activation during the scanning period is measured. If the similarity value between two regions is larger than or equal to a given threshold, then those two regions are assumed to be connected in the resulting functional brain network. Otherwise, those two regions are not connected.

While it may sound intuitive to represent brain activity using a network, it is important to remember that these networks are the result of a complex data transformation process. Therefore, there is a risk that what we observe in a network is partly an artifact of the design choices behind the generation process, and thus not an accurate representation of the underlying system under study. This report is part of an emerging literature focusing on the validity of brain network generation from fMRI data \citep{korhonen2021principles}. In particular, we experimentally quantify the impact of some design choices on various features of the resulting networks.

In Section \ref{sec:pipeline} we review the preprocessing pipeline through which fMRI data is transformed into a network. In Section \ref{sec:problemStatement} we discuss three parameters that mostly affect our understanding about the existence of functional correlations between the brain regions. Specifically, in this section we ask how functional correlation values between regions change based on the change of the value of the three parameters. In Section \ref{sec:proposedModelingMethod} we conclude to model the existence of functional correlations between pairs of the brain's regions as probabilistic edges, not to lose the uncertainty that is inherent in the network generation process. 

\section{Modeling Pipeline} \label{sec:pipeline}

\begin{figure}
\center
    \includegraphics[width=\textwidth]{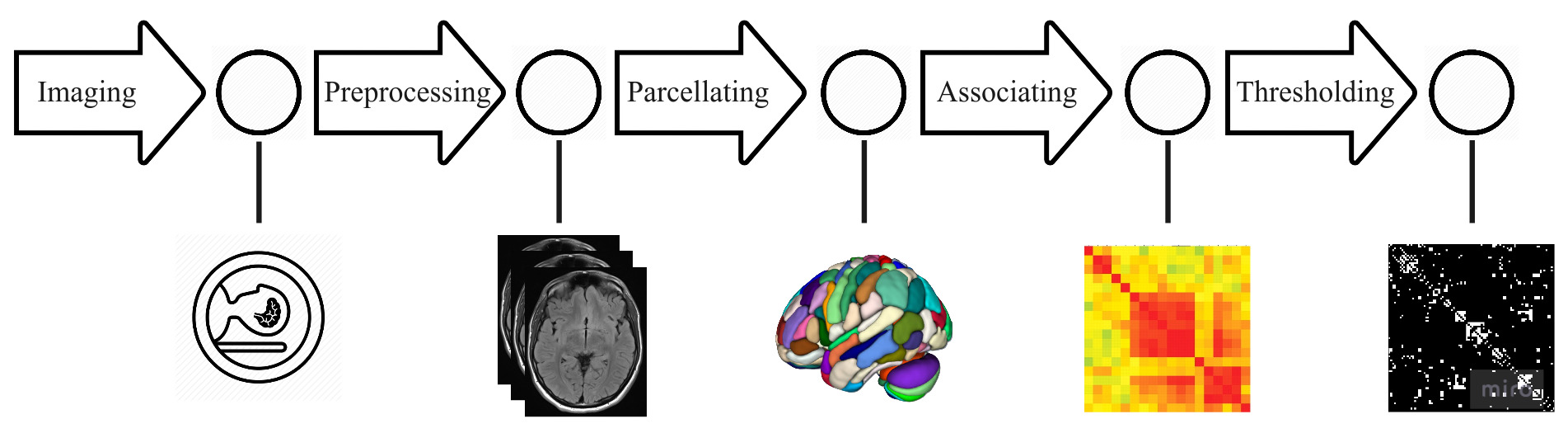}    
\caption{Modeling Pipeline} 
\label{fig:pipeline}
\end{figure}

In this section we review different steps through which a resting state functional brain network is generated (see Figure \ref{fig:pipeline}). The objective is to highlight possible sources of uncertainty that we then study experimentally.

\subsection{Functional Magnetic Resonance Imaging (fMRI)} \label{subsec:fMRI}

fMRI is an imaging method that is used to measure and evaluate interactions between the brain regions by tracking changes in blood flow in the brain. The blood flow in the brain creates some signals which are referred to as Blood-Oxygen-Level-Dependent (BOLD) signals and are detectable by fMRI. Functional imaging can be performed in two manners. First, stimulating the brain regional interactions by giving some task to subjects to be performed during the scanning. Tasks range from simple ones, such as moving a specific finger consecutively, to more complicated tasks such as judging wines based on their smell (aroma) by sommeliers \citep{banks2016structural}. This group of functional imaging are referred to as task-based fMRI. The second way of performing fMRI is called resting state fMRI in which subjects brains are in a task-free state meaning that no specific task is assigned to subjects during the brain scanning. Resting state fMRI has gained attention among researchers as it reveals the spontaneous fluctuations in the brain and therefore it is convenient to explore the changes in the brain functional interactions causing by psychiatric disorders. In this report we focus on resting state fMRI as it is a more active research area. Data from resting state fMRI need to be cleaned from noise and distortions, which is the objective of the second stage of the pipeline: preprocessing.

\subsection{Preprocessing} \label{subsec:preprocessing}

fMRI signals inherently suffer from several noise sources, such as head movement, cardiac/respiratory pulsation or scanner-induced artifacts \citep{arslan2017connectivity}. This leads to major challenges in using and interpreting MRI data. As extracted features in brain networks, such as nodes and edges, are based on processing of MRI data, several preprocessing stages are required to alleviate the effects of noise in MRI signals. Noise removal is more critical in fMRI imaging, since the detected connections between two brain regions are based on the correlation between those regions' time series and adding the same noise to the time series increases the correlation, in turn increasing the chance of incorrectly detecting edges in the brain network (false positives) \citep{arslan2017connectivity,cole2010advances}.

Preprocessing steps can be divided into two groups. First, the group of standard preprocessing steps on which there is a consensus among scholars such as slice timing correction, head motion correction, (magnetic field) distortion correction \citep{ganzetti2016intensity,roy2011intensity}, registration, spatial normalization and spatial smoothing \citep{glasser2013minimal,yang2020constructing}. The second group, includes the preprocessing steps which are controversial and there is no agreement on applying or not applying them. This group includes filtering and global signal regression (GSR).

\textbf{Filtering} - 
BOLD signals are typically low frequency signals. For example, frequency range of BOLD signals in resting state fMRI are usually between 0.01 and 0.1 Hz \citep{arslan2017connectivity,toga2015brain}. The removal of frequencies that are out of the range of interest is called temporal filtering. However, applying temporal filtering on all fMRI datasets is contentious as it has been shown that this preprocessing step increases the variance of correlations of the brain regions and raises false positive rate in detecting connectivity among regions \citep{davey2013filtering}.

\textbf{GSR} - 
Analysis of resting state fMRI signals demonstrates that some components in timeseries are common in all voxels\footnote{A 3-dimensionsal fMRI image is constitutes of units called voxels. Each voxel represents a small cube of brain tissue.}. These components are called global signals \citep{murphy2017towards}. Global signals arising from different sources such as arterial CO2 concentration, blood pressure and the vascular effects \citep{zhu2015vascular}. Hence, global signal removal is desirable since it does not reflect any neural related activity. However, it has been argued that it is also likely that global signals include a neuronal component \citep{fox2009global}. Therefore, keeping or removing the global signals, which is called global signal regression, is also a controversial issue. We refer the interested reader to \citep{li2019global} for a more detailed discussion about the advantages and the disadvantages of GSR.

\begin{figure}
\center
    \includegraphics[width=\textwidth]{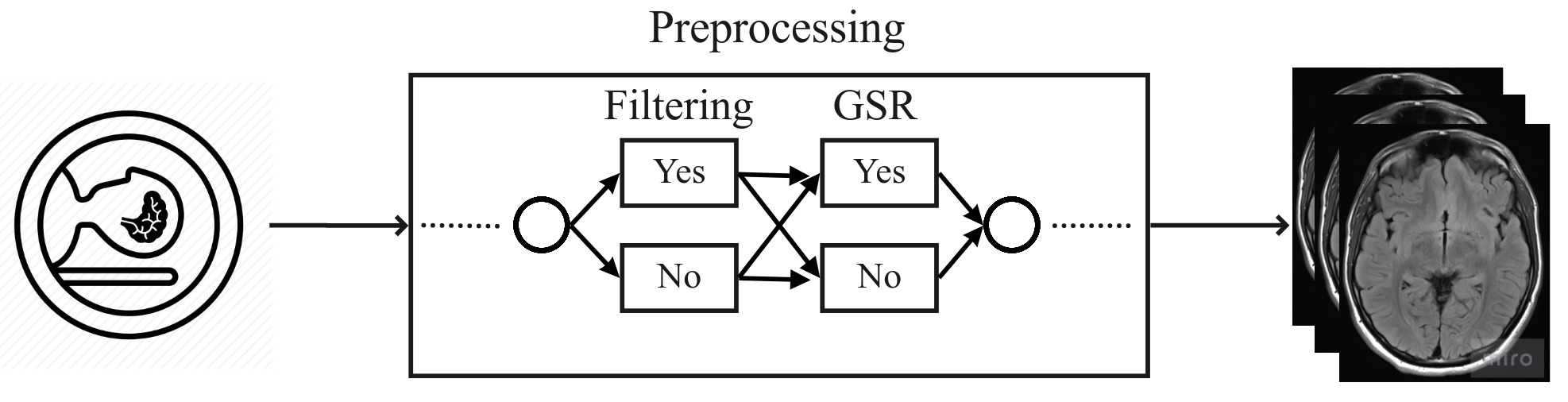}    
\caption{preprocessing critical steps} 
\label{fig:preprocessing-ctitical-steps}
\end{figure}

The resolution of neuroimages with the current available scanning machines is at least 1mm $\times$ 1mm $\times$ 1mm. Thus we need a method to aggregate cubes of voxels based on some criterion to find out the regions that can be seen as a single unit. This processing step, described in the next section, is called parcellating.

\subsection{Parcellating} \label{subsec:parcel}
Parcellation is the process of dividing the brain or part of the brain (e.g., cerebrum) into neuroanatomical regions. A number of parcellation schemes have been developed over the years which are broadly divided into two groups: those schemes in which parcellation is based on fixed structural or anatomical features and those schemes in which regions are aggregated based on functional similarities \citep{de2014graph,termenon2016reliability}. 

Several anatomical parcellation schemes exist, with the number of parcels, called regions of interest (ROIs), typically ranging from 70 to 250 \citep{stanley2013defining}. Standard Automated Anatomical Labeling (AAL) is the most widely used anatomical parcellation scheme, which partitions cerebral cortex into 90 parcels (45 for each hemisphere) \citep{tzourio2002automated}. A modified version of the standard AAL scheme consisting of 89 parcels is used by \cite{termenon2016reliability}. The full AAL parcellation scheme which not only includes cerebral but also cerebellar regions includes 116 parcels \citep{stanley2013defining}. 
There are more fine-grained variations of AAL in which the same parcels in the standard AAL are subdivided into smaller regions such as 459 \citep{alexander2012discovery} and 4320 parcels \citep{fornito2010network}. Another parcellation scheme which is widely used by researchers is the Harvard-Oxford atlas in which the number of ROIs is between 117 \citep{termenon2016reliability} and 300 regions \citep{alexander2012discovery}. The process of subdividing ROIs into smaller regions have been performed by partitioning a parcel into multiple smaller and roughly equal-sized parcels \citep{alexander2012discovery,stanley2013defining}.

Another approach to parcellate the cerebrum and the cerebellum is based on functional similarities between voxels. In this approach, first the similarity between close voxels, in the terms of distance, is calculated and second a clustering algorithm is run to partition the voxels into K clusters. The most widely applied functional parcellations are ICA \citep{calhoun2001method,jafri2008method} and Craddock \citep{craddock2012whole}. The number of parcels varies based on the user-defined K which is an input into the algorithm. Anatomical schemes are mostly used to analyze task-based fMRI and functional similarity schemes are dominantly applied in analysis of resting state fMRI. The parcels or ROIs are the nodes of the brain network resulting out of the preprocessing pipeline.

After selecting and performing the parcellation scheme the next sub-stage is extracting the changes in the intensity of the BOLD signals during the scanning period, called timeseries. The output of this stage, which is a set of ROIs and a set of timeseries corresponding to the ROIs, is passed to the next step, associating, to find the similarity between ROIs based on their timeseries.

\subsection{Associating and Thresholding} \label{subsec:assoc&threshold}
Associating is the forth step of the pipeline which evaluates the similarity between two ROIs' timeseries \citep{de2014graph}. There are multiple methods to find associations between two ROIs such as Pearson correlation coefficient, mutual information \citep{chai2009exploring} and phase coherence \citep{chavez2010functional}. The most applied method is calculating the absolute value of Pearson correlation coefficient of two timeseries which is a value between 0 and 1. The higher correlation value between a pair of ROIs indicates the stronger functional connectivity between those regions. The next step is thresholding which acts as a filtering step, filtering out the correlation values that are lower than a threshold and passing those values that are higher and represent significant functional connectivity. The survived correlations are then set to 1. There has been a long debate about selecting an appropriate threshold value in the literature \citep{de2014graph}. 

The output of this step, which is the final output of the modeling pipeline, is an $N \times N$ matrix with values 1 and 0, and defines the adjacency matrix of a brain network with $N$ nodes.

\section{Problem Statement} \label{sec:problemStatement}

In this report we quantify the effect of different design choices in the process to obtain brain networks from fRMI data. Among all steps of the modeling pipeline, the effect of thresholding on the resulting networks has been studied mostly. Yet, many studies do not account for the uncertainty due to preprocessing and imaging time length. In these studies, the edges between nodes are deterministic, which is the consequence of considering the output of the former steps as certain. In the following three subsections we examine the extent to which imaging time length (Section \ref{subsec:scanning-time}) and the preprocessing choices (Section \ref{subsec:preprocessingChoices}) affect our understanding about the existence of functional correlations/similarity between the brains ROIs. 

To do so, we examine the ABIDE dataset\footnote{Autism Brain Imaging Data Exchange \cite{di2014autism}}. 
The ABIDE dataset includes neuroimaging data from 1112 individuals from which 573 images belong to control individuals and 539 images belong to patients suffering from ASD. The dataset is collected and shared by 16 international imaging sites. All 1112 neuroimaging data are preprocessed by four different methods: the Connectome Computation System (CCS), the Configurable Pipeline for the Analysis of Connectomes (CPAC), the Data Processing Assistant for Resting State fMRI (DPARSF) and the NeuroImaging Analysis Kit (NIAK).

\subsection{Scanning time length} \label{subsec:scanning-time}
Most of the available neuroimaging datasets are a collection of data from different research/imaging centers. Similarly, the ABIDE dataset is a collection of resting state fMRI data from 16 different international research/imaging centers such as UCLA, Stanford and NYU. Since each center has used a different imaging time length, the acquired dataset is not homogeneous in this regard. For example the time length of rs-fMRI images that have been affiliated with CMU (with 315 time slots) is approximately 4.5 times larger than those gathered by OHSU (with 77 time slots) and roughly 2.5 times larger than those reported from Trinity (with 145 slots). Therfore, there has been no consensus between researchers about the required imaging time length, or there can be an implicit assumption that if the imaging time length is long enough, the resulting time series will converge.

\subsection{Preprocessing choices} \label{subsec:preprocessingChoices}
As mentioned in Section \ref{subsec:preprocessing}, among all preprocessing steps, temporal filtering and global signal regression are the most controversial steps because there is no consensus on the usage and usefulness of them. The remaining preprocessing steps such as slice time correction, intensity normalization and motion realignment are uncontroversial. However, there are multiple implementations for each of them. As a result, there is a risk that different implementations lead to different comprehensions of functional similarities among the brain parcels. 

Preprocessing of neuroimaging data in the ABIDE dataset has been performed by four different tools, CCS, CPAC, DPARSF and NIAK. Applying and not applying the two controversial preprocessing steps and being preprocessed in each method makes 16 different combinations. We refer to each combination as a triple $(X, Y, Z)$ where $X \in$ \{filtered, not-filtered\}, $Y \in $ \{global, not-global\} and $Z \in$ \{ccs, cpac, dparsf, niak\}. For example $\mathcal{P}_s =$  (filtered, not-global, ccs) indicates the preprocessing choices for the subject $s$ and shows that we are using the preprocessed input imaging data that has been prepared using the CCS tool, the temporal filtering has been performed over it but the global signals are not removed. As we have 3 parameters for the preprocessing pipeline and one parameter for the imaging time length, we refer to these four parameters as \textit{pipeline parameters vector}.

\begin{mydef}
We refer to the tuple $\{x,y,z,w\}$ as the pipeline parameters vector where $x \in \{$filtered, not-filtered$\}$, $y \in \{$global, not-global$\}$, $z \in \{$ccs, cpac, dparsf, niak$\}$ and $w$ indicates the fraction of the complete timeseries available in the dataset that we use to compute correlations between ROIs.
\end{mydef}

\section{Experiments} \label{subsec:experiments}
In this section we aim at showing the effect of the choices for pipeline parameters vector on our understanding about the existence of functional similarity among the brain's parcels. To inspect the effect of time, we test the correlation values obtained from the first 50\% up to the total reported scanning time length (gradually increasing with 1\%) of the parcels' time series. To investigate the other parameters we examine correlation values obtained from their combinations. 

Our experiments have been designed to address three questions:
\begin{enumerate}
\item How can the pipeline parameters vector affect our understanding about the existence of functional correlation between two parcels? (Section \ref{subsubsec:question1}).
\item How can the pipeline parameters vector affect our understanding about the existence of functional correlation in the whole brain network? (Section \ref{subsubsec:question2}).
\item How can the pipeline parameters vector change our understanding about the difference between the brain networks of an autistic and a control case? (Section \ref{subsubsec:question3}).
\end{enumerate}

\subsection{Functional correlations between two parcels} \label{subsubsec:question1}
\begin{figure}
\center
    \begin{subfigure}{0.35\textwidth}
        \center
        \includegraphics[width=\textwidth]{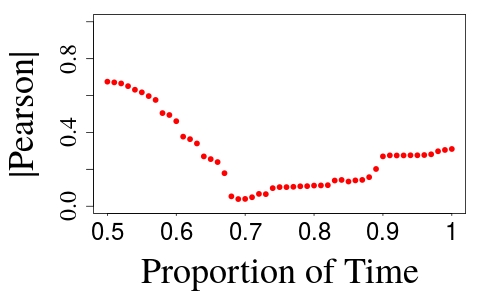}
        \caption{}
        \label{fig:preprocessing_fngcpac_48_79}
    \end{subfigure}
    \begin{subfigure}{0.35\textwidth}
        \center
        \includegraphics[width=\textwidth]{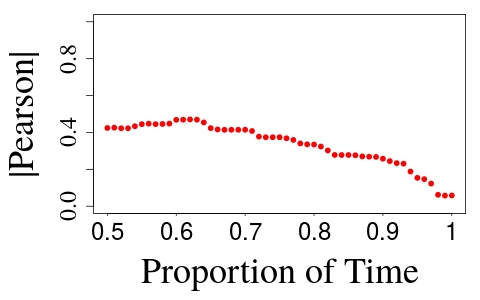}
        \caption{}
        \label{fig:preprocessing_fgccs_1_59}
    \end{subfigure}\\
    \begin{subfigure}{0.35\textwidth}
        \center
        \includegraphics[width=\textwidth]{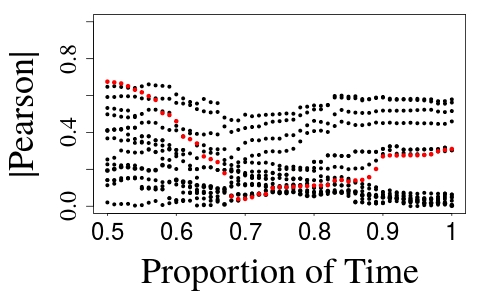}
        \caption{}
        \label{fig:preprocessing_Clatech_51456_all_48_79}
    \end{subfigure}
    \begin{subfigure}{0.35\textwidth}
        \center
        \includegraphics[width=\textwidth]{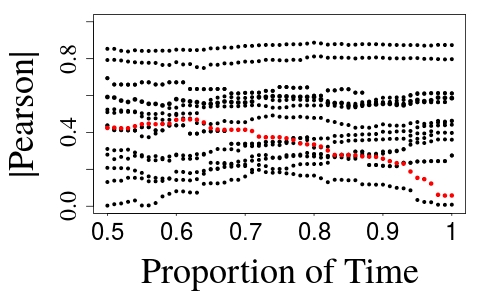}
        \caption{}
        \label{fig:preprocessing_Clatech_51456_all_1_59}
    \end{subfigure}
\caption{\textit{(a)} Correlation coefficients between parcels 48 and 79 in preprocessing triple $(f, ng, cpac)$, \textit{(b)} correlation coefficients between parcels 1 and 59 in preprocessing triple $(f, g, dparsf)$, \textit{(c)} correlation coefficients between parcels 48 and 79 in all 16 preprocessing triples, and \textit{(d)} correlation coefficients between parcels 1 and 59 in all 16 preprocessing triples.} 
\label{fig:time_preprocessing_varies_1_59}
\end{figure}

Figure \ref{fig:preprocessing_fngcpac_48_79} shows the change of functional correlation coefficient between parcels 48 and 79 in a randomly selected subject, \texttt{Caltech\_0051456}, whose time series constitute 145 time slots\footnote{The length of time slots is not explicitly mentioned in the dataset website.}. The subject is a male at age 55.4 who is suffering from autism. The x-axis shows the fraction of the imaging time slots used to compute the correlation. Figure \ref{fig:preprocessing_fngcpac_48_79} shows that if the imaging was finished after 73 slots (half of the whole time series) the functional correlation between parcels 48 and 79 would be larger than 0.6. If the imaging was finished after 100 time slots no functional correlation would be recognized between these two parcels and by increasing the imaging time the functional correlation again increases. This shows that not only there is no unique value of functional correlation for various time lengths, but also there is no linear relation between functional correlation and imaging time length. Figure \ref{fig:preprocessing_fgccs_1_59} also shows that functional correlation between parcels 1 and 59 of the same case has a different pattern. Our experiments over all pairs of parcels show that, 1- functional correlations vary based on the imaging time length, 2- functional correlation does not change linearly as imaging time increases, 3- there is no unique pattern of change of correlation values. In addition, our experiments show that the differences between the maximum and the minimum values of each functional correlation over imaging time are not negligible. For example, this value is larger than 0.2 for more than 94\% of all 4005 pairs of parcels and  is larger than 0.3 for almost 54\% of them. 

Figures \ref{fig:preprocessing_Clatech_51456_all_48_79} and \ref{fig:preprocessing_Clatech_51456_all_1_59} show the change of functional correlation coefficients with all 16 combinations of preprocessing triples. Both figures confirm that functional correlation values do not converge to a specific value neither when we change the preprocessing triples nor when we change the imaging time length.

\begin{figure}
\center
    \begin{subfigure}{0.99\textwidth}
        \center
        \includegraphics[width=\textwidth]{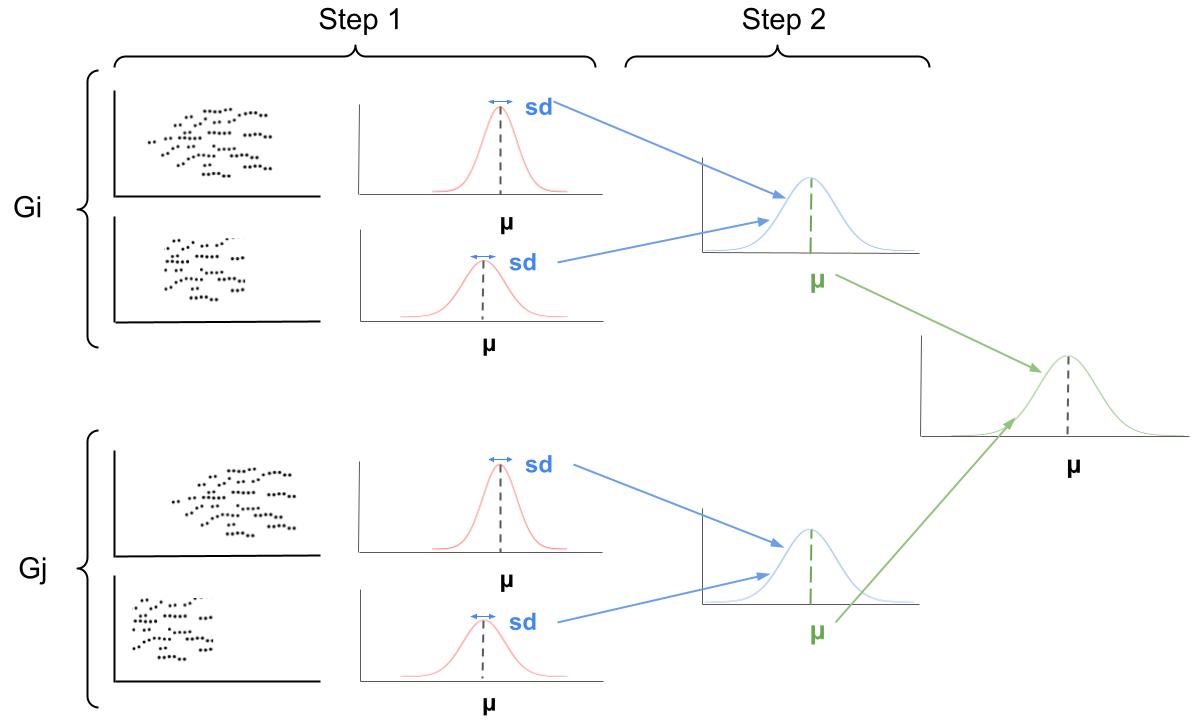}
        \caption{}
        \label{fig:sec3.3.1.workflow}
    \end{subfigure}\\
    \begin{subfigure}{0.32\textwidth}
        \center
        \includegraphics[width=\textwidth]{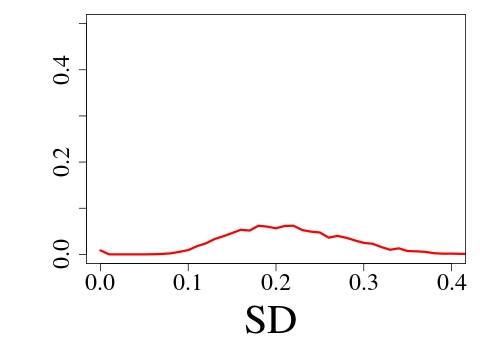}
        \caption{}
        \label{fig:sec3.3.1.sd_oneGraph}
    \end{subfigure}
    \begin{subfigure}{0.32\textwidth}
        \center
        \includegraphics[width=\textwidth]{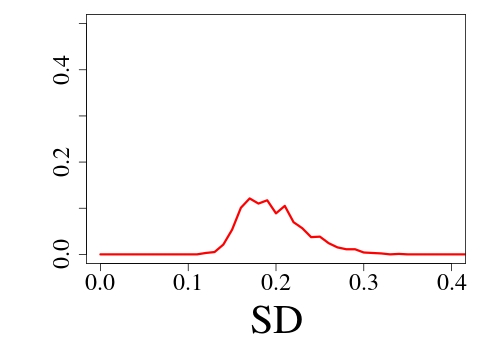}
        \caption{}
        \label{fig:sec3.3.1.mean_of_sd_allGraphs}
    \end{subfigure}
\caption{\textit{(a)} two steps of data processing to highlight the extent to which our understanding about the value of correlation coefficients between pairs of parcels is uncertain. \textit{(b)} The result of step 2 for the case \texttt{Caltech\_51456}, \textit{(c)} the distribution of mean (green $\mu$) over all 1102 brain data.}
\label{fig:sec3.3.1.}
\end{figure}

To represent the variation of functional correlation coefficients on all edges of a brain network, we perform a two step process (see Figure \ref{fig:sec3.3.1.workflow}). First, we generate Pearson correlation coefficients for all pairs of parcels using all (16) preprocessing pipelines and various imaging time lengths. Therefore, for each edge $(u,v)$ we have a probability distribution of correlation coefficients (notated as $p_{uv}(\rho)$) from which we can extract its mean $\mu_{uv}(\rho)$ and standard deviation $\sigma_{uv}(\rho)$. In the second step, we generate the average of standard deviations of all edges $\mu_{g}(\sigma)$, that measures the extent to which correlation coefficients on all edges of the graph are dispersed. A small value of $\mu_{g}(\sigma)$ indicates that the pipeline parameters vector has insignificant impact on our understanding about the existence of functional interactions between ROIs in that brain. On the other hand, if the value of $\mu_{g}(\sigma)$ is not very small then it confirms that the selection of pipeline parameters vector will affect our understandings about the existence of functional interactions. Figure \ref{fig:sec3.3.1.sd_oneGraph} shows this information for the case \texttt{Caltech\_51456}.

Figure \ref{fig:sec3.3.1.mean_of_sd_allGraphs} shows the distribution of $\mu_{g}(\sigma)$ for all 1102 cases in the ABIDE dataset. The figure shows that $\mu_{g}(\sigma)$ for none of the cases in that dataset is significant.

\subsection{Functional correlations among all pairs of parcels} \label{subsubsec:question2}
For the case \texttt{Caltech\_0051456} we generated two brain networks. For both networks the preprocessing triple and the threshold value are $(f,g,ccs)$ and $0.3$ respectively. The imaging time length for the first graph is 70 slots and for the second one is 140. The number of edges in the first and second graphs are 830 and 835 respectively. Among these edges 471 edges are common in both graphs,  359 edges are exclusively available in G1, and 164 edges are exclusively in G2. This shows that with the same choices of parameters in the modeling pipeline but just different imaging time lengths, only 57\% of the edges in G1 and G2 are common. Further investigation on these two graphs shows that Jaccard coefficient of nodes with top-10 degrees in G1 and G2 is 0.35. This coefficient for betweenness is 0 and for closeness is 0.23. Moreover, Jaccard coefficient of the most internal cores\footnote{k-core of graph $G$ is the maximal connected subgraph of $G$ in which all nodes have degree at least k. The most internal core is the k-core with the maximum k.} in G1 and G2 is 0.6. 

Further investigations over graphs resulting from constant preprocessing and threshold choices but varying imaging time length shows that they have inconsistent local as well as global properties.

To represent the effect of the pipeline parameters vector on our understanding about the global properties of the brain networks, we perform experiments to answer the following question: to what extent are two graphs of the same brain but resulting from different pipelines similar? To measure the similarity, we use four metrics. In the first one, we measure Jaccard similarity coefficients of set of edges in both graphs. If it is high (close to 1) this means that the two graphs are similarly modeling the brain and if it is a small value, this means that they are modeling the same graph differently. The second and the third metrics measure the Jaccard similarity coefficients of sets of top-k nodes that are ranked by degree and betweenness centrality respectively. If these values are close to 1, this means that in both graphs the same regions of the brain have been chosen as the areas that have high centrality in the functionality of the brain. The forth metric is the Jaccard similarity between two sets of the most internal cores in two graphs. 

\begin{figure}
\center
    \begin{subfigure}{0.32\textwidth}
        \center
        \includegraphics[width=\textwidth]{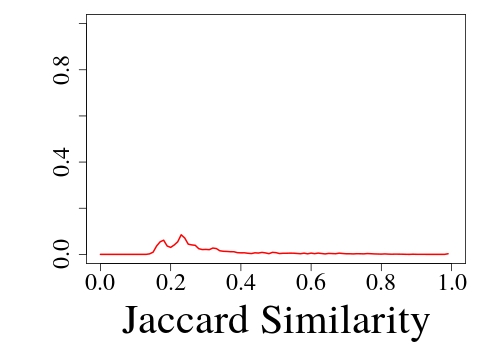}
        \caption{}
        \label{fig:sec3.3.2.pairwiseEdgeStructures_oneBrain}
    \end{subfigure}
    \begin{subfigure}{0.32\textwidth}
        \center
        \includegraphics[width=\textwidth]{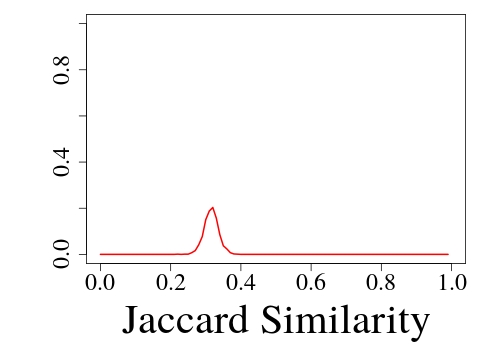}
        \caption{}
        \label{fig:sec3.3.2.pairwiseEdgeStructures_allBrains}
    \end{subfigure}
\caption{\textit{(a)} Jaccard similarity coefficient of the graphs modeling a unique brain \texttt{Caltech\_51456}. \textit{(b)} distribution of mean of Jaccard similarities of all the cases in the dataset.}
\label{fig:sec3.3.2.pairwiseEdgeStructures}
\end{figure}

The timeseries of each brain have been processed with 16 preprocessing pipelines and 6 different imaging time lengths $\{0.5T,0.6T,0.7T,0.8T,0.9T,T\}$, where $T$ is the total time length that is available in the dataset for that brain. Therefore, each brain has been modeled with 96 different graphs. We obtain the Jaccard similarity of edges in each pair of graphs for each brain. Figure \ref{fig:sec3.3.2.pairwiseEdgeStructures_oneBrain} shows this distribution for the case \texttt{Caltech\_51456}. Figure \ref{fig:sec3.3.2.pairwiseEdgeStructures_allBrains} shows the mean of Jaccard similarity for all 1102 cases in the dataset. Both figures show that graph models of unique brains are not similar to each other.

\begin{figure}
\center
    \begin{subfigure}{0.32\textwidth}
        \center
        \includegraphics[width=\textwidth]{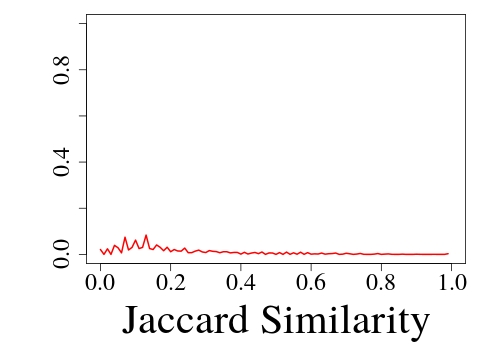}
        \caption{top-20 degree}
        \label{fig:sec3.3.2.pairwise_sim_degree_oneBrains}
    \end{subfigure}
    \begin{subfigure}{0.32\textwidth}
        \center
        \includegraphics[width=\textwidth]{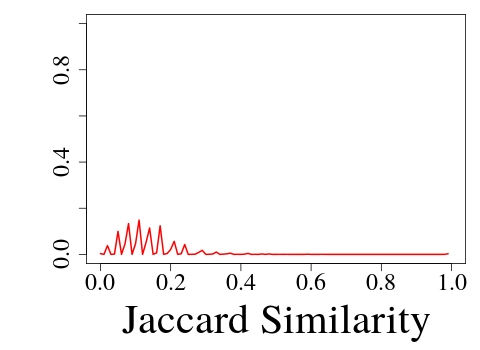}
        \caption{top-20 betweenness}
        \label{fig:sec3.3.2.pairwise_sim_betweenness_oneBrains}
    \end{subfigure}
    \begin{subfigure}{0.32\textwidth}
        \center
        \includegraphics[width=\textwidth]{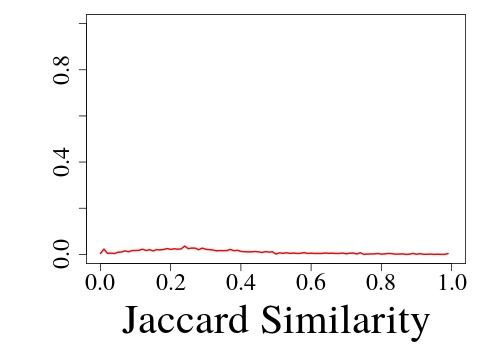}
        \caption{the most internal core}
        \label{fig:sec3.3.2.pairwise_sim_core_oneBrains}
    \end{subfigure}\\
    \begin{subfigure}{0.32\textwidth}
        \center
        \includegraphics[width=\textwidth]{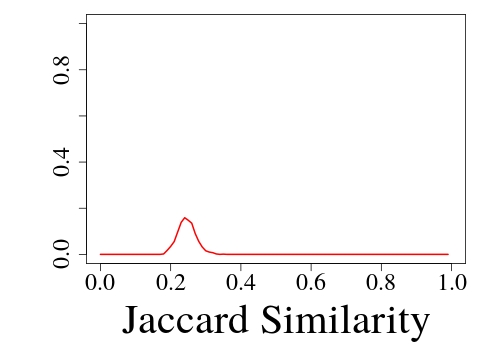}
        \caption{top-20 degree}
        \label{fig:sec3.3.2.sd_of_degree_allBrains}
    \end{subfigure}
    \begin{subfigure}{0.32\textwidth}
        \center
        \includegraphics[width=\textwidth]{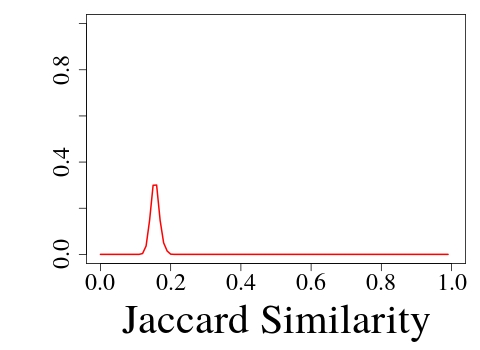}
        \caption{top-20 betweenness}
        \label{fig:sec3.3.2.sd_of_betweenness_allBrains}
    \end{subfigure}
    \begin{subfigure}{0.32\textwidth}
        \center
        \includegraphics[width=\textwidth]{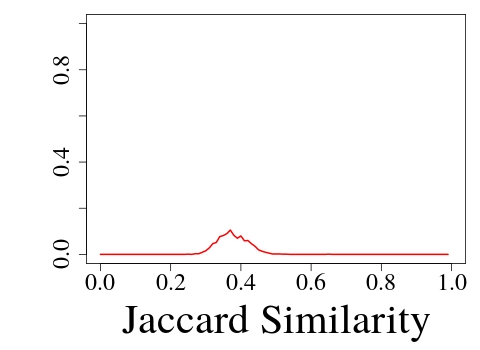}
        \caption{the most internal core}
        \label{fig:sec3.3.2.sd_of_core_allBrains}
    \end{subfigure}
\caption{Similarity between different graphs that model the same brain: \textit{(a)} probability distribution of Jaccard similarity coefficients of top-20 nodes obtained by degree between pairs of graphs modeling the brain \texttt{Caltech\_51456}, \textit{(b)} probability distribution of Jaccard similarity coefficients of top-20 nodes obtained by betweenness between pairs of graphs modeling the brain \texttt{Caltech\_51456} \textit{(c)} probability distribution of Jaccard similarity coefficients of the most internal cores between pairs of graphs modeling the brain \texttt{Caltech\_51456} \textit{(d)} probability distribution of Jaccard similarity coefficients of 20 top degree nodes for all cases in the dataset \textit{(e)} probability distribution of Jaccard similarity coefficients of 20 top betweenness nodes for all cases in the dataset, and \textit{(f)} probability distribution of Jaccard similarity coefficients of the most internal cores for all cases in the dataset.}
\label{fig:sec3.3.2.}
\end{figure}

Figures \ref{fig:sec3.3.2.pairwise_sim_degree_oneBrains} and  \ref{fig:sec3.3.2.pairwise_sim_betweenness_oneBrains} show Jaccard similarity coefficients of top-20 nodes (among 116 nodes) that have been ranked by degree and betweenness centralities respectively for the case \texttt{Caltech\_51456}. Figure \ref{fig:sec3.3.2.pairwise_sim_core_oneBrains} shows the distribution of Jaccard similarity coefficients for the most internal cores in each pair of graphs modeling the brain \texttt{Caltech\_51456}. Figures \ref{fig:sec3.3.2.sd_of_degree_allBrains}, \ref{fig:sec3.3.2.sd_of_betweenness_allBrains} and \ref{fig:sec3.3.2.sd_of_core_allBrains} show the distribution of mean of similarity coefficients for all the brains in the dataset. All figures illustrate low Jaccard similarity coefficients that highlight the uncertainty about the structure of the graphs that model the brains.

\subsection{Graphs of two cases: 1- ASD and 2- Control} \label{subsubsec:question3}
In Sections \ref{subsubsec:question1} and \ref{subsubsec:question2} we investigated the extent to which the modeling pipeline's parameters affect our understanding about the existence of functional correlation between the brain's ROIs in a single case. However, as we mentioned in the Introduction, the second purpose of modeling the brain functionality as a graph is to develop methods to compare healthy brains from unhealthy ones. 

In this section we aim at showing how using different values of modeling pipeline's parameters changes our understanding about the difference between healthy and sick brains. In this experiment we use the data about two cases: \texttt{Caltech\_0051456} as the ASD case and \texttt{Caltech\_0051475} as the control/healthy case. The selected cases are similar in any aspect: 1) the nuroimaging data of both cases have been collected in the same center, \texttt{Caltech}, with the same time length of 145 time slots, 2) both cases are male, 3) both cases are middle-aged, 55.4 and 44.2 respectively, 4) the preprocessing triples and the threshold values for both cases are (filtered, global, ccs) and $0.5$ respectively.

For the healthy case we generate two graphs H1 and H2 corresponding to imaging time length 72 and 145 slots accordingly. For the autistic case we create graphs A1 and A2 corresponding to the same time lengths 72 and 144. To compare the brain graphs of the healthy and the autistic cases, we investigate 1) the graphs' cores, and top-10 nodes of graphs chosen by 2) degree, 3) betweenness, and 4) normalized closeness (harmonic).

\begin{figure}
\center
    \begin{subfigure}{0.7\textwidth}
        \center
        \includegraphics[width=\textwidth]{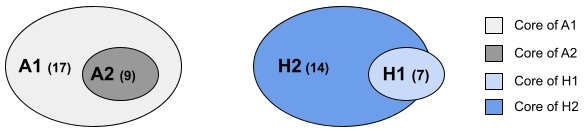}
        \caption{}
        \label{fig:cores_1_fgccs}
    \end{subfigure}\\
    \begin{subfigure}{0.7\textwidth}
        \center
        \includegraphics[width=\textwidth]{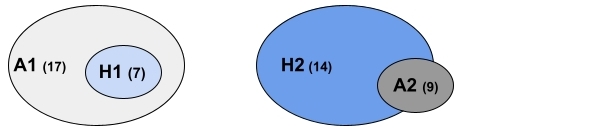}
        \caption{}
        \label{fig:cores_2_fgccs}
    \end{subfigure}
\caption{}
\label{fig:cores_fgccs}
\end{figure}

Figure \ref{fig:cores_1_fgccs} shows that the most internal core of the autistic case shrinks as the imaging time length increases from 72 to 145 slots. On the other hand, the most internal core of the healthy case expands. Moreover, Figure \ref{fig:cores_2_fgccs} shows that if we use the imaging time length 72 slots, the core of the healthy case is a subset of the core of the autistic case, however when we use the imaging time length 145, the cores of the healthy and autistic case are getting separated (6 nodes are in common). 

\begin{figure}
    \centering
    \includegraphics[width = 0.45\textwidth]{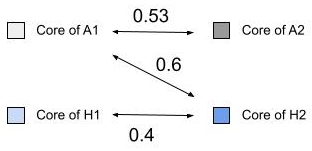}
    \caption{Jaccard similarity coefficients}
    \label{fig:CoreJaccard}
\end{figure}

In addition, the Jaccard coefficients of the most internal cores in A1 and A2 is 0.53 and for H1 and H2 is 0.4, which shows some degree of similarity between the cores. However this coefficient between the most internal cores of A1 and H2 is 0.6 which is larger than the other two (see Figure \ref{fig:CoreJaccard}).

As a result this experiment shows that our understanding about the difference between the brain networks of healthy and autistic cases changes when only the imaging time length differs and other parameters are constant. By fixing the imaging time length and changing the preprocessing parameters, we observe the same diverging results.

\begin{figure}
\center
    \begin{subfigure}{0.45\textwidth}
        \center
        \includegraphics[width=\textwidth]{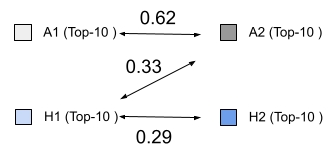}
        \caption{degree}
        \label{fig:top10_degree}
    \end{subfigure}
    \begin{subfigure}{0.45\textwidth}
        \center
        \includegraphics[width=\textwidth]{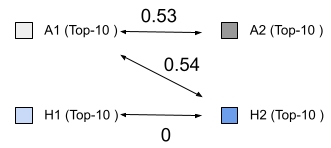}
        \caption{closeness}
        \label{fig:top10_closeness}
    \end{subfigure}
\caption{Similarities computed using the Jaccard coefficient}
\label{fig:top10_centralities}
\end{figure}

To compare the difference between the top-10 ROIs in the autistic and the control brain we use degree, betweenness and normalized closeness (harmonic). Figure \ref{fig:top10_degree} shows the Jaccard coefficients of the top-10 degrees nodes in A1 and A2 is 0.62 and for H1 and H2 is 0.29. However this coefficient between the top-10 degrees nodes in A2 and H1 is 0.33 which is larger than the similarity between H1 and H2. The same experiments for the top-10 closeness nodes shows that the similarity value between A1 and H2 is larger than A1 and A2 as well as H1 and H2.

The previous example only highlights the effect of imaging time length on our understanding about the distinctive features between an autistic and a control case. However, we intend to show effects of different values for pipeline parameters vector on the features that make distinction between control and autistic groups. To do so, we extract two simple levels of features: (1) node-level features, and (2) edge-level features. To extract node-level features, for unique pipeline parameters vector we obtain the set of nodes that are among the top-K nodes in at least in 95\% of the control (autistic) cases. The top-K nodes are chosen based on three different measures: i) degree, ii) betweenness, and iii) the core numbers. To extract edge-level features, for each pipeline parameters vector we obtain the set of edges that are available in at least in 95\% of the control (autistic) cases.

For node-level features, our analysis shows that pipeline parameters vectors, the set of nodes that are among the K = \{10, 20, 30\} top degree/betweenness nodes in at least 95\% of the control (autistic) cases is empty. This shows that we could not find any similarity regarding the top nodes (using degree and betweenness) in each group no matter which pipeline has been used. In other words, we could not find a node-wise feature that is present in at least 95\% of the cases in each group. Our results for the most internal cores are similar.

\begin{figure}
\center
    \begin{subfigure}{0.32\textwidth}
        \center
        \includegraphics[width=\textwidth]{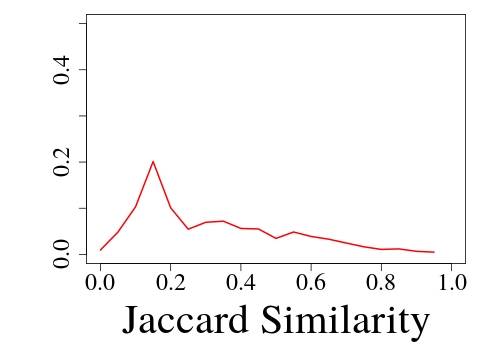}
        \caption{Autistic}
        \label{fig:AllPs_edges_in_A95}
    \end{subfigure}
    \begin{subfigure}{0.32\textwidth}
        \center
        \includegraphics[width=\textwidth]{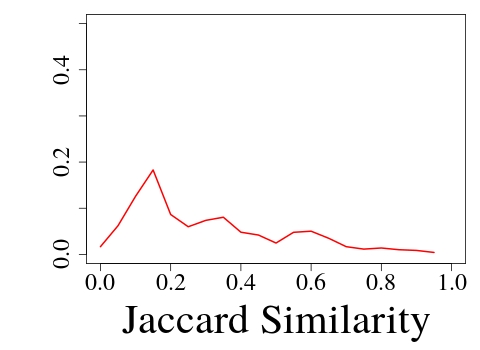}
        \caption{Control}
        \label{fig:AllPs_edges_in_C95}
    \end{subfigure}
    \begin{subfigure}{0.32\textwidth}
        \center
        \includegraphics[width=\textwidth]{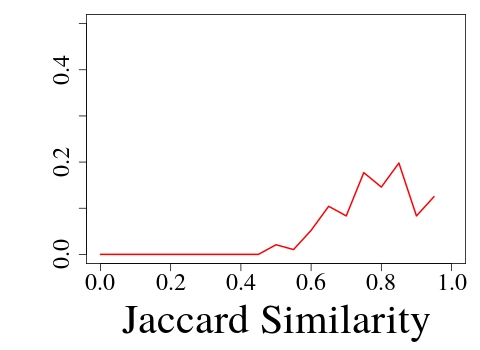}
        \caption{Autistic and Control}
        \label{fig:eachP_edges_in_A95_C95}
    \end{subfigure}
\caption{The similarity between sets of 95\% common edges resulting from various preprocessing choices}
\label{fig:Ps_and_edge_sim_95}
\end{figure}

\begin{figure}
\center
    \begin{subfigure}{0.32\textwidth}
        \center
        \includegraphics[width=\textwidth]{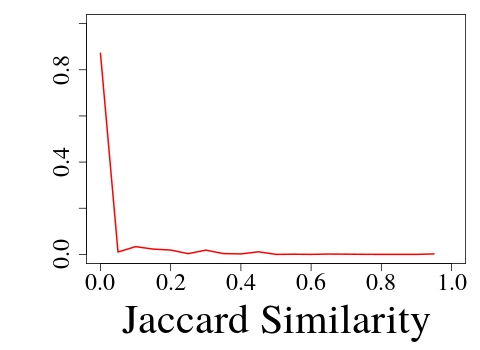}
        \caption{Autistic but not Control}
        \label{fig:P1_Jaccard_top20_degree_A_and_C}
    \end{subfigure}
    \begin{subfigure}{0.32\textwidth}
        \center
        \includegraphics[width=\textwidth]{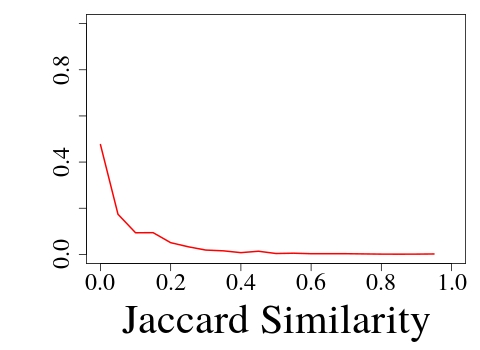}
        \caption{Control but not Autistic}
        \label{fig:P1_Jaccard_top20_degree_C_and_A}
    \end{subfigure}
\caption{Unique sets of 95\% common edges that are exclusively available in the first group: \textit{(a)} exclusively available in the autistic group, \textit{(b)} exclusively available in the autistic group.}
\label{fig:P1_and_All_Ps_Jaccard_top20_degree}
\end{figure}

While node-level analysis does not give us useful information, edge level analysis is more informative. In edge level analysis, for each pipeline parameters vector, we extract the set of edges that are available in at least 95\% of the autistic (control) graphs. Then we calculate the Jaccard similarity coefficients of pairs of edge sets to show the extent to which different preprocessing choices identify similar or different sets of edges as the feature of the autistic (control) group. Figures \ref{fig:AllPs_edges_in_A95} and \ref{fig:AllPs_edges_in_C95} show this information for the autistic and control groups respectively. While similarity between pairs of sets in each group is not that much high (highlights the uncertainty in information about the set of edges that are available in at least in 95\% of the cases in each group), similarity between the sets of edges in two different groups is higher (see Figure \ref{fig:eachP_edges_in_A95_C95}). More precisely, while the similarity between sets of 95\% common edges resulting from various preprocessing choices for each group is not very high (the property that we would like to be high), the similarity between sets of 95\% common edges between the control group and the autistic group is high (the property that we would like to be low). Figure \ref{fig:P1_and_All_Ps_Jaccard_top20_degree} shows the similarity between sets of 95\% common edges that are exclusively available in the autistic group (Figure \ref{fig:P1_Jaccard_top20_degree_A_and_C}) and the control group (Figure \ref{fig:P1_Jaccard_top20_degree_C_and_A}). The similarities are very low.

\section{Discussion and Outlook}
\label{sec:proposedModelingMethod}

In this report we showed that choosing different preprocessing parameters and imaging time lengths can lead to completely different networks modeling the brain data. 
As a consequence, we argue that producing a single deterministic network as the end of the pipeline only provides a partial representation of the information available in the fMRI data.

A possible direction that we believe is worth exploring to address this problem is to use probabilistic network models instead of deterministic ones to represent uncertainty in functional interactions between brains regions in fMRI data. 
A probabilistic network is a network in which edges are associated with an independent existence probability. More precisely, a probabilistic network is notated as $\mathcal{G}(V,E,p)$ where $V$ is the set of nodes, $E$ is the set of edges and $p: E \to (0,1]$ is a function which assigns a probability to each edge $e \in E$.

As we showed in Section \ref{subsec:experiments} the functional correlation coefficient between the timeseries of two parcels varies throughout imaging time. This suggests that a probability distribution, and not a single number, can be a better representation for functional correlation coefficients between parcels. Therefore, for each pair of parcels we have a probability distribution of functional correlation coefficients. A probability distribution of functional correlation coefficients is also assigned to each pair of parcels
by \cite{zanin2021uncertainty}\footnote{In this work probability distributions are calculated based on Bayesian methods}, so that networks can be then generated by sampling these distributions.

\begin{figure}
\center
    \begin{subfigure}{0.32\textwidth}
        \center
        \includegraphics[width=\textwidth]{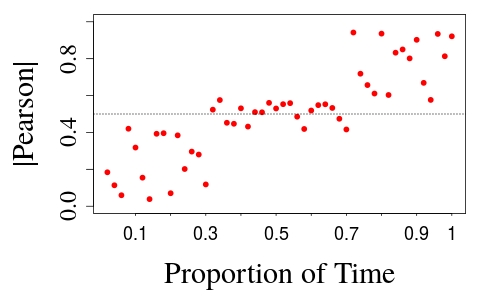}
        \caption{}
        \label{fig:imaginary_parcels_1}
    \end{subfigure}
    \begin{subfigure}{0.32\textwidth}
        \center
        \includegraphics[width=\textwidth]{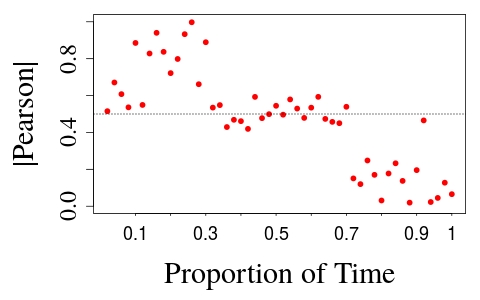}
        \caption{Control but not Autistic}
        \label{fig:imaginary_parcels_2}
    \end{subfigure}
\caption{Correlation coefficients between two pairs of parcels}
\label{fig:imaginary_parcels}
\end{figure}

\begin{figure}
\center
    \begin{subfigure}{0.32\textwidth}
        \center
        \includegraphics[width=\textwidth]{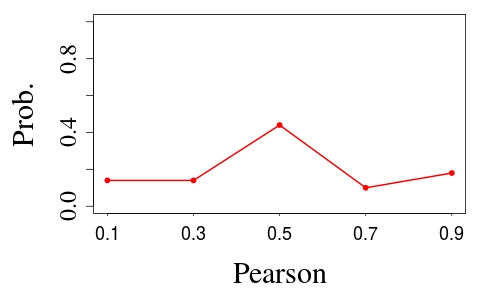}
        \caption{}
        \label{fig:imaginary_parcels_dist_1}
    \end{subfigure}
    \begin{subfigure}{0.32\textwidth}
        \center
        \includegraphics[width=\textwidth]{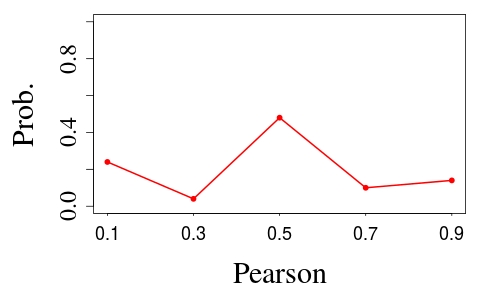}
        \caption{Control but not Autistic}
        \label{fig:imaginary_parcels_dist_2}
    \end{subfigure}
\caption{Probability distribution of correlation coefficients between the two pairs of parcels shown in Figure \ref{fig:imaginary_parcels}.}
\label{fig:imaginary_parcels_dist}
\end{figure}

Calculating the probability distribution of functional correlation coefficients is useful when we are looking at each pair of parcels independently. However, if we intend to study the brain as a system (macro analysis) then using probability distributions of functional correlation coefficients without considering time dependency between the distributions can be misleading. Take for example Figures \ref{fig:imaginary_parcels_1} and \ref{fig:imaginary_parcels_2}, showing the variation of the functional correlation coefficient between two pairs of parcels over time. Although they are showing two different behaviors, their probability distributions of functional correlation coefficients are 
very similar (Figure \ref{fig:imaginary_parcels_dist}). Sampling such a distribution leads to  instances in which both functional correlations are high. However, by considering time, we can see that when functional correlation between the first pair of parcels is high, it is low for the second pair and vice versa. 

One way to avoid sampling unrealistic instances of macro-scale functional correlations between parcels, can be to first sample the time and second to calculate functional correlation coefficients at sampled time points. As a result, given a preprocessing pipeline and $N$, we can sample randomly the imaging time length $N$ times and generate a weighted graph out of each sample. At the third step, by thresholding each weighted network we have $N$ simple networks that all together represent the functional interactions between parcels of a subject's brain. Given $N$ simple networks of a subject's brain we can generate a probabilistic network in which an edge probability is determined by the relative frequency of samples in which the edge exists. 

While this discussion shows that probabilistic networks can be generated as an output of the pipeline, maintaining more information from the originating fMRI data, additional research is needed to verify whether and how this enriched output can be exploited in the subsequent network analysis phases.

\bibliographystyle{unsrtnat}
\bibliography{references}

\end{document}